\documentclass[11pt,showpacs,preprintnumbers,amsmath,amssymb,prd,nofootinbib,superscriptaddress]{revtex4-2}

\usepackage{dcolumn}
\usepackage{bm}
\usepackage{ifpdf}
\usepackage{hyperref}
\usepackage{bm}
\usepackage{xcolor,color,graphicx,graphics,physics}
\usepackage[spanish,english]{babel}
\usepackage[latin1]{inputenc}
\usepackage[OT1]{fontenc}
\usepackage{latexsym,amssymb,amsmath,amsfonts, slashed}
\usepackage{makeidx}
\usepackage{epsfig,subfigure}
\usepackage{natbib}
\usepackage{epstopdf}
\usepackage{mathrsfs}
\usepackage{hyperref}
\hypersetup{colorlinks=true, linkcolor=blue, citecolor=blue, urlcolor=blue}
\usepackage{enumerate}
\usepackage{cancel}
\usepackage{braket}
\usepackage{fixmath}
\usepackage{comment}

\everymath{\displaystyle}
\usepackage{graphicx}

\usepackage{amsmath}
\usepackage{amssymb}
\usepackage{graphicx}
\usepackage{xcolor}
\newcommand{\pa}{\partial}

\newcommand{\vk}{{\bf k}}
\newcommand{\bea}{\begin{eqnarray}}
\newcommand{\eea}{\end{eqnarray}}
\newcommand{\orcid}[1]{\href{https://orcid.org/#1}{\includegraphics[width=10pt]{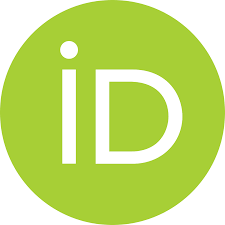}}}

\begin{document}

\title{Thermal and spatial confinement effects in Podolsky electrodynamics}

\author{L. H. A. R. Ferreira \orcid{0000-0002-4384-2545}}
\email{luiz.ferreira@fisica.ufmt.br}
\affiliation{Programa de P\'{o}s-Gradua\c{c}\~{a}o em F\'{\i}sica, Instituto de F\'{\i}sica,\\ 
Universidade Federal de Mato Grosso, Cuiab\'{a}, Brasil}

\author{A. F. Santos \orcid{0000-0002-2505-5273}}
\email{alesandroferreira@fisica.ufmt.br}
\affiliation{Programa de P\'{o}s-Gradua\c{c}\~{a}o em F\'{\i}sica, Instituto de F\'{\i}sica,\\ 
Universidade Federal de Mato Grosso, Cuiab\'{a}, Brasil}

\begin{abstract}

In this work, Podolsky theory, a second-order, Lorentz- and gauge-invariant extension of classical electrodynamics, is considered. The effects of Podolsky's modification on fundamental phenomena such as the Stefan-Boltzmann law and the Casimir effect for the electromagnetic field are investigated. The Thermo Field Dynamics (TFD) formalism is employed to describe quantum fields at finite temperature and under spatial confinement through its topological structure.
			
\end{abstract}

\maketitle

\section{Introduction}
	
The classical theory of electromagnetism is fundamentally described by Maxwell electrodynamics, from which a vast range of scientific and technological advancements has emerged. This theory is grounded in Maxwell's four equations, which inherently ensure charge conservation through the continuity equation and reflect the underlying symmetries of the electromagnetic field. However, despite its robustness, Maxwell's theory exhibits certain limitations, particularly in the presence of divergences at small scales. In response to these challenges, an extension known as Podolsky electrodynamics was proposed by B. Podolsky in 1942 \cite{p1,p2}. This generalized framework introduces a second-order derivative term into the Lagrangian, while preserving both Lorentz and gauge invariance. One of the key features of this modification is the emergence of a massive sector for the photon, achieved without violating gauge symmetry. The additional higher-derivative term serves to regularize the theory at high frequencies, effectively acting as a natural cutoff. A notable example of this regularizing effect appears in the electrostatic potential: whereas Maxwell's theory predicts a divergence as $r \to 0$, Podolsky electrodynamics yields a finite, non-zero value at the origin \cite{p3}. Podolsky electrodynamics has been investigated in a variety of physical contexts, particularly in the formulation of plasma models based on generalized electrodynamics with higher-order derivatives. The presence of an additional massive mode, alongside the usual massless photon, modifies both the propagation properties of the gauge field and the effective interactions it mediates across different energy scales. In the finite-temperature regime, such modifications directly affect the collective properties of the plasma. In~\cite{ref1}, it is shown that although the general structure of electrostatic screening is preserved, the Debye mass becomes dependent not only on temperature, as in conventional QED, but also on the parameter associated with the Podolsky term. Within the Matsubara formalism, an explicit one-loop calculation of the polarization tensor reveals that the static limit of the longitudinal sector incorporates contributions from the additional massive mode, leading to quantitative corrections to the exponential suppression of the electrostatic potential.

At the level of effective descriptions, these modifications are reflected in the formulation of Abelian plasma theories coupled to quantized electromagnetic fields with Podolsky corrections~\cite{ref2}. The resulting Hamiltonian structure exhibits modified photon dispersion relations and new propagation modes. Upon integrating out the photonic degrees of freedom, an effective plasma Hamiltonian emerges in which particle-mediated mode--mode couplings arise, along with collective effects such as mode mixing, photon condensation, and non-Markovian behavior in finite systems. The interplay between thermal screening and ultraviolet regularization leads to more stable intermediate-range interactions at high energies, suggesting potential relevance for confined plasmas, solid-state systems, and laser--plasma interactions. Furthermore, the incorporation of gauge-preserving mass-generation mechanisms further extends the theoretical scope of the model. In~\cite{ref3}, a formulation combining Podolsky electrodynamics with a Stueckelberg-type mechanism is developed, introducing a compensating scalar field that preserves the $U(1)$ gauge symmetry while dynamically linking the infrared regime, associated with a Proca-type mass, to the ultraviolet regime controlled by the higher-derivative scale. Finally, a consistent microscopic foundation for relativistic kinetic plasma theory is provided in~\cite{ref4}, where it is argued that the Vlasov--Maxwell equations should be derived from the regularized Bopp--Land\'e--Thomas--Podolsky electrodynamics. In the continuum limit, this approach leads to a Vlasov--Maxwell--BLTP-type system including $1/N$ corrections and radiation-reaction effects, while recovering the standard Vlasov--Maxwell system in the singular limit $\kappa \to \infty$. Taken together, these results indicate that higher--derivative gauge theories not only provide improved ultraviolet behavior and better mathematical control, but also lead to potentially measurable physical consequences in the collective properties and effective dynamics of interacting systems across multiple energy scales.

In this work, Podolsky electrodynamics is studied within the framework of Thermo Field Dynamics (TFD). Thermal and size effects are then investigated by exploring the topological structure inherent to this formalism. Quantum field theory at finite temperature can be formulated through several approaches, such as the imaginary-time formalism \cite{matsubara}, the closed-time path formalism \cite{sch,kel}, and the Thermo Field Dynamics (TFD) formalism, the latter of which is adopted in this work. Proposed by H. Umezawa and Y. Takahashi in 1975, TFD is characterized by an elegant structure that enables the description of quantum systems under thermal or spatial confinement \cite{thermofield}. In this framework, an equivalence is established between the statistical average of an operator and its vacuum expectation value, leading to the definition of a thermal Hilbert space $\mathcal{H}_T$, constructed as the tensor product $\mathcal{H} \otimes \tilde{\mathcal{H}}$ between the original Hilbert space and its tilde-conjugate. As a consequence, the degrees of freedom of the system are doubled. A key element of this approach is the Bogoliubov transformation, which introduces temperature or spatial dependence by rotating the original variables. Moreover, TFD can be interpreted as a topological field theory, with a topology given by $\Gamma^d_D = \mathbb{S}^1 \times \mathbb{R}^{D-d}$, where $D$ denotes the spacetime dimensions and $d$ represents the number of compactified coordinates. This topological structure allows various physical effects to be examined: for instance, thermal effects are introduced by compactifying the time dimension into a circle of circumference $\beta$ \cite{tqft,ume}, while spatial confinement is modeled by compactifying one or more spatial dimensions, all leading to a toroidal topology.

To investigate the implications of Podolsky electrodynamics within the framework of the Thermo Field Dynamics (TFD) formalism, three distinct topological configurations are considered. In the first case, thermal effects are examined through the derivation of the Stefan-Boltzmann law. The second application involves a topological structure in which spatial confinement is introduced, leading to the computation of the Casimir effect. Originally proposed by H. Casimir in 1948, this phenomenon describes the emergence of an attractive force between two conducting parallel plates as a consequence of boundary conditions or topological constraints \cite{c1, c2, c3, c4}. The third scenario explores a combined compactification of temporal and spatial dimensions, allowing for the simultaneous analysis of thermal and confinement effects -- namely, the Casimir effect at finite temperature. Accordingly, the primary goal of this manuscript is to determine the modifications introduced by Podolsky's theory to classical electrodynamics by employing the TFD formalism.
	
This paper is organized as follows. In Section \ref{2}, the Thermo Field Dynamics (TFD) formalism is introduced. Section \ref{3} presents Podolsky's Lagrangian and the construction of the coordinate-space propagator. In Section \ref{4}, the energy-momentum tensor for Podolsky electrodynamics is derived within the TFD framework. Section \ref{5} is dedicated to specific applications: Subsection \ref{51} addresses thermal effects by compactifying the temporal coordinate; Subsection \ref{52} explores size effects through the compactification of the spatial $z$-axis; and Subsection \ref{53} examines the combined effects of thermal and spatial confinement by compactifying both time and $z$ coordinates. Finally, Section \ref{6} provides concluding remarks.

\section{Thermo Field Dynamics formalism}\label{2}
	
The Thermo Field Dynamics (TFD) formalism provides a powerful framework for describing quantum systems at finite temperature. It is formulated as a real-time thermal quantum field theory, based on the equivalence between the statistical average of an operator and its vacuum expectation value. This approach allows one to retain the full temporal structure of the system. Within this formulation, the thermal Hilbert space emerges as the tensor product of the original Hilbert space and an auxiliary, or tilde, Hilbert space: ${\cal H}_T = {\cal H} \otimes \tilde{\cal H}$. This construction leads to a doubling of the degrees of freedom \cite{thermofield}.

Another essential element of the TFD formalism is the Bogoliubov transformation. This transformation performs a rotation between the tilde and non-tilde variables, introducing a dependence on a parameter $\alpha = (\alpha_0, \alpha_1, ..., \alpha_D)$, known as the compactification parameter. The term ``compactification'' highlights a topological feature of the TFD approach: the ability to wrap spacetime dimensions into a toroidal topology. This structure enables the simultaneous description of both thermal (temperature-related) and spatial (finite-size) effects.

Given two arbitrary operators ${\cal O}$ and $\tilde{\cal O}$ acting on the Hilbert spaces ${\cal H}$ and $\tilde{\cal H}$, respectively, the application of the Bogoliubov transformation leads to
	\begin{align}
		\left( \begin{array}{cc} {\cal O}(k, \alpha)  \\\xi \tilde {\cal O}^\dagger(k, \alpha) \end{array} \right)={\cal U}(\alpha)\left( \begin{array}{cc} {\cal O}(k)  \\ \xi\tilde {\cal O}^\dagger(k) \end{array} \right),
	\end{align}
where ${\cal U}(\alpha)$ is the Bogoliubov transformation defined as
	\begin{align}
		{\cal U}(\alpha)=\left( \begin{array}{cc} u(\alpha) & -v(\alpha) \\
			\xi v(\alpha) & u(\alpha) \end{array} \right)
	\end{align}
with $u^2(\alpha)+\xi w^2(\alpha)=1$. Here, $\xi=-1(+1)$ for bosons(fermions).

To illustrate an important application, let us consider the construction of the photon propagator within this framework. Due to the Bogoliubov transformation, the resulting propagator acquires a dependence on the parameter $\alpha$. Consequently, the photon propagator takes the form
	\begin{align}
		iD_{\mu\nu}(x-x';\alpha)&=\braket{0,\tilde{0}|\tau[A_\mu^a(x;\alpha)A_\nu^b(x';\alpha)]0,\tilde{0}},\nonumber\\
		&=\eta_{\mu\nu}G_0^{(ab)}(x-x';\alpha).
	\end{align}
Here, $a, b = 1, 2$, $\tau$ denotes the time-ordering operator, and $\eta_{\mu\nu}$ is the Minkowski metric. The gauge field is defined such that
$A^a_\mu(x; \alpha) = {\cal U}(\alpha), A^a_\mu(x), {\cal U}^{-1}(\alpha)$,
where ${\cal U}(\alpha)$ is the Bogoliubov transformation operator. The quantity $G_0^{(ab)}$ corresponds to the massless scalar field propagator, defined as
	\begin{align}
		G_0^{(ab)}(x-x';\alpha)=i\int \frac{d^4k}{(2\pi)^4}e^{-ik(x-x')}G_0^{(ab)}(k;\alpha),
	\end{align}
	where $G_0^{(ab)}(k;\alpha)={\cal U}(\alpha)G_0^{(ab)}(k){\cal U}^{-1}(\alpha)$ with
	\begin{align}
		G_0^{(ab)}(k)=\left( \begin{array}{cc} G_0(k) & 0 \\
			0 & \xi G^*_0(k) \end{array} \right),
	\end{align}
and $G_0(k)=-1/k^2$. The quantities of interest are associated with the non-tilde variables; therefore, the physical Green's function is given by the component $a = b = 1$, that is,
	\begin{align}
		G_0^{(11)}(k;\alpha)=G_0(k)+\xi v^2(k;\alpha)[G^*_0(k)-G_0(k)],
	\end{align}
where $v^2(k;\alpha)$ is given by
	\begin{align}
		v^2(k;\alpha)=\sum_{s=1}^d\sum_{\lbrace\sigma_s\rbrace}2^{s-1}\sum_{l_{\sigma_1},...,l_{\sigma_s}=1}^\infty(-\xi)^{s+\sum_{r=1}^sl_{\sigma_r}}\,\exp\left[{-\sum_{j=1}^s\alpha_{\sigma_j} l_{\sigma_j} k^{\sigma_j}}\right].
	\end{align}
This expression corresponds to the generalized Bogoliubov transformation, where $d$ is the number of compactified dimensions, ${\sigma_s}$ denotes the set of all combinations with $s$ elements, and $k$ represents the four-momentum \cite{tqft,ume}.
	
In the next section, we will discuss Podolsky's Lagrangian, with the main objective of constructing the propagator associated with this theory.
	
\section{The Podolsky Lagrangian and Its Propagator in Position Space}\label{3}

This section presents the Podolsky propagator in position space, which will be a key tool in the next section, where the TFD formalism is applied. The starting point is Podolsky electrodynamics, described by the following Lagrangian
	\begin{align}
		{\cal L}=-\frac{1}{4}F^{\mu\nu}F_{\mu\nu}+\frac{a^2}{2}\partial_\alpha F^{\alpha\beta}\partial_\lambda F^\lambda_{\;\beta},\label{8}
	\end{align}
where the electromagnetic field tensor is defined as $F^{\mu\nu} = \partial^\mu A^\nu - \partial^\nu A^\mu$. 

The generalization of classical electrodynamics in Podolsky theory arises from the inclusion of a higher-order derivative term involving the gauge field $A^\mu$. An important feature of this extension is the Podolsky parameter $a$, which has the inverse dimension of length. This parameter effectively introduces a massive sector for the photon, with the associated mass given by the relation $a = \hbar / mc$ \cite{p1,p2,p3}. In this context, the photon acquires a second mode of propagation, one massless and one massive, which modifies the behavior of the electromagnetic field at short distances.

To obtain the propagator in momentum space, the Lagrangian must be rewritten in the following form
	\begin{align}
		{\cal L}&=\frac{1}{2}\big[A_\nu\Box A^\nu-A^\mu\pa_\nu\pa_\mu+\pa_\mu(A_\nu\pa^\mu A^\nu)-\pa_\nu(A^\mu\pa_\mu A^\nu)\big]+\frac{a^2}{2}\big[A^\beta\Box\Box A_\beta+\pa_\alpha(\pa^\alpha A^\beta\Box A_\beta)\nonumber\\&-\pa^\alpha(A^\beta\pa_\alpha\Box A_\beta)-A^\beta\Box\pa_\lambda\pa_\beta A^\lambda-2\pa_\alpha(\pa^\alpha A^\beta\pa_\lambda\pa_\beta A^\lambda)+2\pa^\alpha(A^\beta\pa_\alpha\pa_\lambda\pa_\beta A^\lambda)\nonumber\\&+A^\alpha\Box\pa_\alpha\pa_\lambda A^\lambda+\pa_\alpha(\pa^\beta A^\alpha\pa_\lambda\pa_\beta A^\lambda)-\pa^\beta(A^\alpha\pa_\alpha\pa_\lambda\pa_\beta A^\lambda)\big].
	\end{align}
Then, by substituting into the action, the following expression is obtained
\begin{align}
		S&=\int dx^4\frac{1}{2}\big[A_\nu\Box A^\nu-A^\mu\pa_\nu\pa_\mu A^\nu+a^2A^\beta\Box\Box A_\beta-a^2A^\beta\Box\pa_\lambda\pa_\beta A^\lambda\big]+\frac{1}{2}\int dx^4\big[\pa_\mu(A_\nu\pa^\mu A^\nu)\nonumber\\&-\pa_\nu(A^\mu\pa_\mu A^\nu)\big]+\frac{a^2}{2}\int dx^4\big[ \pa_\alpha(\pa^\alpha A^\beta\Box A_\beta)-\pa^\alpha(A^\beta\pa_\alpha\Box A_\beta)-2\pa_\alpha(\pa^\alpha A^\beta\pa_\lambda\pa_\beta A^\lambda)\nonumber\\&+2\pa^\alpha(A^\beta\pa_\alpha\pa_\lambda\pa_\beta A^\lambda)+A^\alpha\Box\pa_\alpha\pa_\lambda A^\lambda+\pa_\alpha(\pa^\beta A^\alpha\pa_\lambda\pa_\beta A^\lambda)-\pa^\beta(A^\alpha\pa_\alpha\pa_\lambda\pa_\beta A^\lambda)\big].
	\end{align}
By applying the divergence theorem, the second and third integrals can be neglected. Additionally, an arbitrary gauge\footnote{Indeed, the gauge-fixing term is introduced into the Lagrangian, so that the Podolsky Lagrangian becomes ${\cal L}=-\frac{1}{4}F^{\mu\nu}F_{\mu\nu}+\frac{a^2}{2}\partial_\alpha F^{\alpha\beta}\partial_\lambda F^\lambda_{\;\beta}-\frac{1}{2\zeta}(\partial_\mu A^\mu)^2$.} will be fixed to facilitate the derivation of the propagator. Consequently, the action takes the form
	\begin{align}
		S=\int dx^4 A^\mu\big[\Box g_{\mu\nu}-\pa_\nu\pa_\mu+a^2g_{\mu\nu}\Box\Box-a^2\Box\pa_\mu\pa_\nu+\frac{1}{\zeta}\pa_\mu\pa_\nu\big]A^\nu.
	\end{align}

The propagator can be obtained by identifying the kernel of the action and applying the following identity
	\begin{align}
		\left(\Box g_{\mu\nu}-\pa_\nu\pa_\mu+a^2g_{\mu\nu}\Box\Box-a^2\Box\pa_\mu\pa_\nu+\frac{1}{\zeta}\pa_\mu\pa_\nu\right)\Delta^{\nu\sigma}_F(x-x')=i\delta^\sigma_{\;\mu}\delta^4(x-x').
	\end{align}
By considering the Fourier transform of the propagator $\Delta^{\mu\nu}_F(x - x')$,
	\begin{align}
		\Delta^{\mu\nu}_F(x-x')=\int\frac{dk^4}{(2\pi)^4}\Delta^{\mu\nu}_F(k)e^{-ik\cdot(x-x')},
	\end{align}
we obtain the following equation in momentum space
	\begin{align}
		\left(-k^2g_{\mu\nu}+k_\mu k_\nu+a^2g_{\mu\nu}k^4-a^2k^2k_\mu k_\nu-\frac{1}{\zeta}k_\mu k_\nu\right)\Delta^{\nu\rho}_F(k)=i\delta^\rho_{\;\mu}.
	\end{align}
An ansatz of the form
	\begin{align}
		\Delta^{\nu\rho}_F(k)={\cal A}g^{\nu\rho}+{\cal B}k^\nu k^\rho
	\end{align}
is assumed. By substituting this into the equation for the propagator, the following solution is obtained
	\begin{align}
		{\cal A}=\frac{1}{a^2k^4-k^2},\quad\quad{\cal B}=\frac{\zeta-a^2k^2\zeta-1}{k^2(a^2k^4-k^2)}.
	\end{align}
Thus, the propagator in momentum space is given by
	\begin{align}
		\Delta^{\mu\nu}_F(k)=-\frac{1}{k^2(a^2k^2-1)}\left[\theta^{\mu\nu}+\zeta(1-a^2k^2)\omega^{\mu\nu}\right],
	\end{align}
where $\theta^{\mu\nu}=g^{\mu\nu}-\omega^{\mu\nu}$ and $\omega^{\mu\nu}=k^\mu k^\nu/k^2$ are the transverse and longitudinal projectors, respectively, and $\zeta$ is an arbitrary gauge parameter. By considering the Feynman gauge, $\zeta = 1$, the following expression can be written
	\begin{align}
		D^{\mu\nu}(k)=\frac{1}{k^2(a^2k^2-1)}\left[g^{\mu\nu}-a^2k^\mu k^\nu\right].
	\end{align}
From the above expression, it can be seen that the propagator exhibits a singularity with two poles. This gives rise to two independent equations, written as
\begin{align}
	\vk^2=\omega^2\label{20}
\end{align}
and
\begin{align}
	\omega^2&=\vk^2+\frac{1}{a^2}.\label{21}
\end{align}
These correspond to the dispersion relations of Podolsky's theory. Eq. \eqref{20} represents the standard relation for the massless photon sector, while Eq. \eqref{21} describes the massive sector introduced by the Podolsky modification. Both relations are essential for constructing the propagator in position space. It should be noted that all calculations were performed in natural units, where $\hbar=c=k_B=1$.

To obtain the propagator, a Fourier transformation will be performed by writing
	\begin{align}
		D^{\mu\nu}(x-x')=(g^{\mu\nu}-a^2k^\mu k^\nu)\int\frac{d^4k}{(2\pi)^4}\frac{e^{-ik(x-x')}}{k^2(a^2k^2-1)}.
	\end{align}
The term in the integral can be rewritten as
	\begin{align*}
		\frac{1}{k^2(a^2k^2-1)}=-\frac{1}{k^2}+\frac{a^2}{a^2k^2-1}.
	\end{align*}
Then, it can be written as
	\begin{align}
		D^{\mu\nu}(x-x')=(g^{\mu\nu}+a^2\partial^\mu\partial^\nu)\left[-\underbrace{\int\frac{d^4k}{(2\pi)^4}\frac{e^{-ik(x-x')}}{k^2}}_I+\underbrace{a^2\int\frac{d^4k}{(2\pi)^2}\frac{e^{-ik(x-x')}}{a^2k^2-1}}_{II}\right]\label{23},
	\end{align}
where the relation $k^\mu = -i \partial^\mu$ has been used. The integral $I$ is known and yields
	\begin{align}
		I=\frac{i}{4\pi^2}\frac{1}{(x-x')^2}\label{24}.
	\end{align}
The integral $II$ must be evaluated in order to obtain the propagator. Accordingly, the calculation is carried out by
	\begin{align}
		II=a^2\int\frac{d^4k}{(2\pi)^4}\frac{e^{-ik(x-x')}}{a^2(k_0^2-\vec{k}^2-\frac{1}{a^2})}=\int\frac{d^3k}{(2\pi)^3}e^{i\vec{k}\cdot(\vec{x}-\vec{x}')}\int\frac{dk}{(2\pi)}\frac{e^{-ik_0(x_0-x_0')}}{k_0^2-\omega^2}.
	\end{align}
By applying the residue theorem and expressing the integral in spherical polar coordinates, the following form is obtained
	\begin{align}
		II&=-\frac{i}{(2\pi)^3}\int_{0}^{\infty}dk\int_{0}^{\pi}\sin\theta d\theta\int_{0}^{2\pi}d\phi\frac{k^2}{2\omega}e^{-i\omega t+ikr\cos\theta}\nonumber\\
		&=-\frac{1}{8\pi^2r}\partial_r\int dk\frac{e^{-\omega t+ikr}}{\omega},
	\end{align}
where	$t=x_0-x_0'$ and $r=\vec{x}-\vec{x}'$. The following definitions are introduced
	\begin{align}
		\omega=\frac{\cosh\eta}{a}\quad\quad \mbox{and}\quad\quad k=\frac{\sinh\eta}{a},
	\end{align}
which satisfy the dispersion relation \eqref{21}. Thus, it follows that
	\begin{align}
		II=\frac{i}{8\pi^2 r}\partial_r\int_{-\infty}^{\infty} d\eta \exp[-\frac{i}{a}(t\cosh\eta-r\sinh\eta)].
	\end{align}
To proceed, three cases for the spacetime separation can be considered: timelike, spacelike, and lightlike. In this work, the focus is restricted to the spacelike separation, where $(x - x')^2 < 0$, or equivalently, $|x_0 - x_0'| < r$. After performing the integral, it is found that
	\begin{align}
		II&=-\frac{im}{4\pi^2}\frac{K_1\left(m\sqrt{-(x-x')^2}\right)}{\sqrt{-(x-x')^2}}\label{34},
	\end{align}
where natural units ($\hbar = c = 1$) are used, the mass is defined as $m = 1/a$, and $K_\nu(z)$ denotes the modified Bessel function.

Then, by substituting the results from \eqref{24} and \eqref{34} into \eqref{23}, the propagator is expressed as
	\begin{align}
		D^{\mu\nu}(x-x')=-\left(g^{\mu\nu}+\frac{\partial^\mu\partial^\nu}{m}\right)\left[\frac{i}{4\pi^2}\frac{1}{(x-x')^2}+\frac{im}{4\pi^2}\frac{K_1\left(m\sqrt{-(x-x')^2}\right)}{\sqrt{-(x-x')^2}}\right].\label{35}
	\end{align}
It is important to emphasize that this propagator is consistent, as it correctly reduces to the standard Maxwell propagator in the appropriate limit $a \to 0$, which corresponds to the mass parameter $m \to \infty$.

The restriction to spacelike separation is physically motivated by the fact that the quantities under consideration correspond to static vacuum observables. In configurations involving time-independent boundary conditions, such as those associated with the Casimir effect, the relevant vacuum expectation values of the field operators are evaluated at equal times, $x^0 = x'^0$. Under this condition, the invariant interval reduces to $(x-x')^2 = -|\mathbf{x}-\mathbf{x}'|^2 < 0$, corresponding to spacelike separation. Therefore, timelike and lightlike separations do not contribute to the equal-time vacuum correlations associated with the static configuration \cite{spacelike, weinberg, plates}.

Now, the energy-momentum tensor for Podolsky electrodynamics can be written to explore applications within the TFD formalism and its associated topological structure.
	
\section{Podoslky energy-momentum tensor}\label{4}

In this section, the Lagrangian (\ref{8}) describing Podolsky electrodynamics is considered to derive the corresponding energy-momentum tensor. First, the tensor is obtained in its standard form, and then the tools of TFD are applied to express it as a function of the topological parameter.

Using the standard definition, the symmetric energy-momentum tensor associated with Podolsky electrodynamics is given by \cite{b1}
\begin{eqnarray}
T^{\mu\nu}&=-F^{\mu\alpha}F^\nu_{\;\;\alpha}+\frac{1}{4}g^{\mu\nu}F_{\alpha\beta}F^{\alpha\beta}+a^2{\bigg(}-\frac{1}{2}g^{\mu\nu}\partial_\alpha F^{\alpha\beta}\partial_\gamma F^\gamma_{\;\;\beta}-F^{\mu\alpha}\Box F^\nu_{\;\;\alpha}\nonumber\\
		&-F^{\nu\alpha}\Box F^\mu_{\;\;\alpha}-F^{\mu\alpha}\partial_\alpha\partial_\beta F^{\beta\nu}-F^{\nu\alpha}\partial_\alpha\partial_\beta F^{\beta\mu}+\partial_\tau F^{\tau\mu}\partial_\gamma F^{\gamma\nu}{\bigg)}.
\end{eqnarray}
	
It is important to note that the product of fields evaluated at the same spacetime point leads to divergences when computing the vacuum expectation value of the energy-momentum tensor. To address this issue, we consider the time-ordered product of the fields evaluated at distinct spacetime points, written as
	\begin{eqnarray}
		T^{\mu\nu}(x)&=\lim_{x'\to x}\tau{\Bigg\{}-F^{\mu\alpha}(x)F^\nu_{\;\;\alpha}(x')+\frac{1}{4}g^{\mu\nu}F_{\alpha\beta}(x)F^{\alpha\beta}(x')+a^2{\bigg(}-\frac{1}{2}g^{\mu\nu}\partial_\alpha F^{\alpha\beta}(x)\partial_\gamma F^\gamma_{\;\;\beta}(x')\nonumber\\&-F^{\mu\alpha}(x)\Box F^\nu_{\;\;\alpha}(x')-F^{\nu\alpha}(x)\Box F^\mu_{\;\;\alpha}(x')-F^{\mu\alpha}(x)\partial_\alpha\partial_\beta F^{\beta\nu}(x')-F^{\nu\alpha}(x)\partial_\alpha\partial_\beta F^{\beta\mu}(x')\nonumber\\&+\partial_\tau F^{\tau\mu}(x)\partial_\gamma F^{\gamma\nu}(x'){\bigg)}{\Bigg\}}\label{49}.
	\end{eqnarray}
By applying the time-ordering operator and evaluating all terms, the energy-momentum tensor can be rewritten as
	\begin{eqnarray}
		T^{\mu\nu}(x)&=\lim_{x'\to x}{\Bigg\{}{\bigg[}-\Gamma^{\mu\nu,\lambda\rho,\alpha}_{\quad\quad\;\;\alpha
		}(x,x')+\frac{1}{4}g^{\mu\nu}\Gamma_{\alpha\beta,}^{\quad\lambda\rho,\alpha\beta}(x,x')+a^2{\bigg(}-\frac{1}{2}g^{\mu\nu}\Pi^{\alpha\gamma,\quad\;\beta\quad\lambda\rho}_{\quad\alpha\gamma,\quad\beta,}(x,x')\nonumber\\&-\Theta^{\mu\gamma,\;\;\;\nu,\alpha\;\;\;\lambda\rho}_{\quad\;\gamma\quad\;\alpha,}(x,x')-\Theta^{\nu\gamma,\;\;\;\mu,\alpha\;\;\;\lambda\rho}_{\quad\;\gamma\quad\;\alpha,}(x,x')-\Sigma^{\mu\quad\beta,\alpha\nu,\lambda\rho}_{\;\;\alpha,\beta}(x,x')-\Sigma^{\nu\quad\beta,\alpha\mu,\lambda\rho}_{\;\;\alpha,\beta}(x,x')\nonumber\\&+\Pi^{\tau\gamma,\quad\mu\nu,\lambda\rho}_{\quad\tau\gamma,}(x,x'){\bigg)}{\bigg]}\tau\left[A_\lambda(x)A_\rho(x')\right]-I^{\mu\alpha,\nu}_{\quad\;\;\alpha}(x,x')+\frac{1}{4}g^{\mu\nu}I_{\alpha\beta,}^{\quad\alpha\beta}(x,x')+a^2\nonumber\\&\times{\bigg(}-\frac{1}{2}g^{\mu\nu}I^{\alpha\quad\beta,\gamma}_{\;\;\alpha,\beta\quad\gamma}(x,x')-I^{\mu\alpha,\;\;\gamma,\nu}_{\quad\gamma\quad\alpha}(x,x')-I^{\nu\alpha,\;\;\gamma,\mu}_{\quad\gamma\quad\alpha}(x,x')-I^{\mu\alpha,\quad\beta\nu}_{\quad\alpha\beta}(x,x')\nonumber\\&-I^{\nu\alpha,\quad\beta\mu}_{\quad\alpha\beta}(x,x')+I_{\tau\quad\gamma,}^{\;\;\tau,\mu\;\;\gamma\nu}(x,x'){\bigg)}{\Bigg\}},
	\end{eqnarray}
where have been defined
	\begin{align}
		\Gamma^{\mu\nu,\lambda\rho,\alpha}_{\quad\quad\;\;\alpha}(x;x')&=(g^{\lambda\rho}\partial^\mu-g^{\mu\lambda}\partial^\alpha)(g^\rho_{\;\;\alpha}\partial'^\nu-g^{\nu\rho}\partial'_\alpha),\\
		\Gamma_{\alpha\beta,}^{\quad\lambda\rho,\alpha\beta}(x;x')&=(g^\lambda_{\;\;\beta}\partial_\alpha-g^\lambda_{\;\;\alpha}\partial_\beta)(g^{\beta\rho}\partial'^\alpha-g^{\alpha\rho}\partial'^\beta),\\
		\Pi^{\alpha\gamma,\quad\;\beta\quad\lambda\rho}_{\quad\alpha\gamma,\quad\beta,}(x,x')&=(g^{\beta\lambda}\Box-g^{\alpha\lambda}\partial_\alpha\partial^\beta)(g^\rho_{\;\;\beta}\Box'-g^{\gamma\rho}\partial'_\gamma\partial'_\beta),\\
		\Theta^{\mu\gamma,\;\;\;\nu,\alpha\;\;\;\lambda\rho}_{\quad\;\gamma\quad\;\alpha,}(x,x')&=(g^{\alpha\lambda}\partial^\mu-g^{\mu\lambda}\partial^\alpha)(g_\alpha^{\;\;\rho}\Box'\partial'^\nu-g^{\nu\rho}\Box'\partial'_\alpha),\\
		\Theta^{\nu\gamma,\;\;\;\mu,\alpha\;\;\;\lambda\rho}_{\quad\;\gamma\quad\;\alpha,}(x,x')&=(g^{\alpha\lambda}\partial^\nu-g^{\nu\lambda}\partial^\alpha)(g_\alpha^{\;\;\rho}\Box'\partial'^\mu-g^{\mu\rho}\Box'\partial'_\alpha),\\
		\Sigma^{\mu\quad\beta,\alpha\nu,\lambda\rho}_{\;\;\alpha,\beta}(x,x')&=(g^{\alpha\lambda}\partial^\mu-g^{\mu\lambda}\partial^\alpha)(g^{\nu\rho}\partial'_\alpha\Box'-g^{\beta\rho}\partial'_\alpha\partial'_\beta\partial'^\nu),\\
		\Sigma^{\nu\quad\beta,\alpha\mu,\lambda\rho}_{\;\;\alpha,\beta}(x,x')&=(g^{\alpha\lambda}\partial^\nu-g^{\nu\lambda}\partial^\alpha)(g^{\mu\rho}\partial'_\alpha\Box'-g^{\beta\rho}\partial'_\alpha\partial'_\beta\partial'^\mu),\\
		\Pi^{\tau\gamma,\quad\mu\nu,\lambda\rho}_{\quad\tau\gamma,}(x,x')&=(g^{\mu\lambda}\Box-g^{\tau\lambda}\partial_\tau\partial^\mu)(g^{\nu\rho}\Box'-g^{\gamma\rho}\partial'_\gamma\partial'^\nu), 
	\end{align}
and $I^{\mu\nu,\rho\lambda,\sigma\zeta}(x,x')$ represents the full set of commutation relations between the field components and their derivatives.

By taking the vacuum expectation value of the energy-momentum tensor and applying the elements of the TFD formalism, the topological effects -- encoded through the Bogoliubov transformation and the compactification parameter -- can be introduced. As a result, we obtain
\begin{eqnarray}
\ev{T^{\mu\nu}(x;\alpha)}=\ev{T^{\mu\nu}(x)}{0(\alpha)}.
\end{eqnarray}
As a consequence, the photon propagator naturally arises and is defined through the thermal vacuum expectation value as $\braket{0(\alpha)|\tau[A^\mu(x)A^\nu(x')]|0(\alpha)}=iD^{\mu\nu}(x-x'; \alpha)$.

To proceed, a renormalization procedure must be carried out. In the Casimir and thermal contexts, the vacuum expectation value of the energy--momentum tensor is obtained from derivatives of the two-point function in the coincident limit $x' \to x$. The divergences arising in this limit are related to the local short-distance structure of the Green function and are independent of the presence of boundaries or temperature. A renormalized quantity is thus defined by subtracting the corresponding Minkowski vacuum contribution,
\[
\langle T_{\mu\nu} \rangle_{\text{ren}} =
\langle T_{\mu\nu} \rangle_{\text{BC/T}} -
\langle T_{\mu\nu} \rangle_{0}.
\]
This subtraction removes the state-independent local singular structure of the Green function, which includes contributions associated with timelike and lightlike propagation. As a result, the finite physical contribution is determined by the modification of the equal-time spatial correlations of the quantum field induced by the boundary conditions or temperature.

In this context, we adopt the Casimir prescription, where the finite energy-momentum tensor is obtained by taking the difference
	\begin{align}
		{\cal T}^{\mu\nu(ab)}(x;\alpha)=\braket{T^{\mu\nu(ab)}(x;\alpha)}-\braket{T^{\mu\nu(ab)}(x)}.
	\end{align}
As a result of this renormalization procedure, the finite energy-momentum tensor is obtained as
	\begin{align}
		{\cal T}^{\mu\nu(ab)}(x;\alpha)&=-i\lim_{x'\to x}{\Bigg\{}\Gamma^{\mu\nu,\lambda\rho,\alpha}_{\quad\quad\;\;\alpha
		}(x,x')-\frac{1}{4}g^{\mu\nu}\Gamma_{\alpha\beta,}^{\quad\lambda\rho,\alpha\beta}(x,x')-a^2{\bigg(}-\frac{1}{2}g^{\mu\nu}\Pi^{\alpha\gamma,\quad\;\beta\quad\lambda\rho}_{\quad\alpha\gamma,\quad\beta,}(x,x')\nonumber\\&-\Theta^{\mu\gamma,\;\;\;\nu,\alpha\;\;\;\lambda\rho}_{\quad\;\gamma\quad\;\alpha,}(x,x')-\Theta^{\nu\gamma,\;\;\;\mu,\alpha\;\;\;\lambda\rho}_{\quad\;\gamma\quad\;\alpha,}(x,x')-\Sigma^{\mu\quad\beta,\alpha\nu,\lambda\rho}_{\;\;\alpha,\beta}(x,x')-\Sigma^{\nu\quad\beta,\alpha\mu,\lambda\rho}_{\;\;\alpha,\beta}(x,x')\nonumber\\&+\Pi^{\tau\gamma,\quad\mu\nu,\lambda\rho}_{\quad\tau\gamma,}(x,x'){\bigg)}{\Bigg\}}\left(g^{\lambda\rho}+a^2\partial^\lambda\partial^\rho\right)\overline{G}_0^{(ab)P}(x-x';\alpha)\label{64},
	\end{align}
where the subtracted propagator
	\begin{align}
		\overline{G}_0^{(ab)P}(x-x';\alpha)=G^{(ab)P}_0(x-x;\alpha)-G^{(ab)P}_0(x-x'),
	\end{align}
is defined in terms of the Podolsky propagator in position space as
	\begin{align}
		G_0^P(x-x')=\frac{i}{4\pi^2}\frac{1}{(x-x')^2}+\frac{im}{4\pi^2}\frac{K_1\left(m\sqrt{-(x-x')^2}\right)}{\sqrt{-(x-x')^2}}.
	\end{align}
	
In Eq. \eqref{64}, the term $\left(g^{\lambda\rho} + a^2 \partial^\lambda \partial^\rho\right)$ multiplies the derivative operators. By carrying out this operation explicitly, the energy-momentum tensor takes the final form
	\begin{align}
		{\cal T}^{\mu\nu(ab)}(x;\alpha)&=-i\lim_{x'\to x}{\Bigg\{}2\partial^\mu\partial'^\nu-\frac{1}{2}g^{\mu\nu}g_{\beta\alpha}\partial^\beta\partial'^\alpha-\frac{1}{m^2}{\bigg(}-\frac{1}{2}g^{\mu\nu}g_{\gamma\rho}g_{\alpha\beta}\partial^\gamma\partial^\beta\partial'^\rho\partial'^\alpha\nonumber\\&-2g_{\gamma\rho}\partial^\mu\partial'^\rho\partial'^\gamma\partial'^\nu-2g_{\gamma\rho}\partial^\nu\partial'^\rho\partial'^\gamma\partial^\mu-2g_{\alpha\beta}\partial^\alpha\partial'^\beta\partial'^\mu\partial'^\nu-g_{\gamma\rho}\partial^\rho\partial^\gamma\partial'^\mu\partial'^\nu\nonumber\\&-g_{\gamma\rho}\partial^\mu\partial^\nu\partial'^\rho\partial'^\gamma+g_{\gamma\rho}\partial^\mu\partial^\gamma\partial'^\rho\partial'^\nu{\bigg)}{\Bigg\}}\overline{G}_0^{(ab)P}(x-x';\alpha).\label{67}
	\end{align}

With the full expression of the energy-momentum tensor now obtained as a function of the topological parameter $\alpha$, we are in a position to explore its applications. In the next section, specific components of the tensor will be computed to investigate the thermal and spatial compactification effects within the Podolsky theory. These analyses will provide insights into how topology and finite-temperature environments modify the behavior of the field.

\section{Applications}\label{5}

To analyze the thermal and spatial confinement effects in Podolsky's electrodynamics, we now turn to the topological framework provided by the TFD formalism. In this context, three distinct topologies will be examined: (i) $\Gamma^1_4 = \mathbb{S}^1 \times \mathbb{R}^3$, characterized by $\alpha = (\beta, 0, 0, 0)$, where compactification occurs along the time axis, accounting for finite temperature; (ii) the same topology $\Gamma^1_4$ but with $\alpha = (0, 0, 0, i2d)$, representing spatial compactification along the $z$-axis; (iii) $\Gamma^2_4 = \mathbb{S}^1 \times \mathbb{S}^1 \times \mathbb{R}^2$, where both the time and $z$ axes are compactified, incorporating simultaneous thermal and spatial effects \cite{tqft,ume}.

\subsection{Thermal effects}\label{51}

To account for thermal effects, the compactification parameter is chosen as $\alpha = (\beta, 0, 0, 0)$. This choice allows for the calculation of the Stefan-Boltzmann law within the framework of Podolsky's electrodynamics. In this thermal scenario, the corresponding Bogoliubov transformation is given by
	\begin{align}
		v^2(\beta)=\sum_{l_0=1}^{\infty}e^{-\beta k^0l_0}
	\end{align}
	and the Green function is
	\begin{align}
		\overline{G}_0^P(x-x';\beta)=2\sum_{l_0=1}^{\infty}G_0^P(x-x'-i\beta l_0n_0).
	\end{align}
	
To derive the Stefan-Boltzmann law within the context of Podolsky's electrodynamics, we consider the component $\mu = \nu = 0$ of  Eq. \eqref{67}, leading to
	\begin{align}
		{\cal T}^{00(11)}(\beta)=\frac{1}{\pi^2}\sum_{l_0=1}^{\infty}\left\{\frac{6}{\beta^4l_0^4}+\frac{9m^2}{2\beta^2l_0^2}K_2(m\beta l_0)+\frac{3m^3}{2\beta l_0}K_1(m\beta l_0)\right\}\label{70}.
	\end{align}
This expression represents the generalized Stefan-Boltzmann law for the electromagnetic field, now modified by the presence of the Podolsky mass parameter $m$.
	
\begin{figure}[!ht]
		\includegraphics[width=8cm]{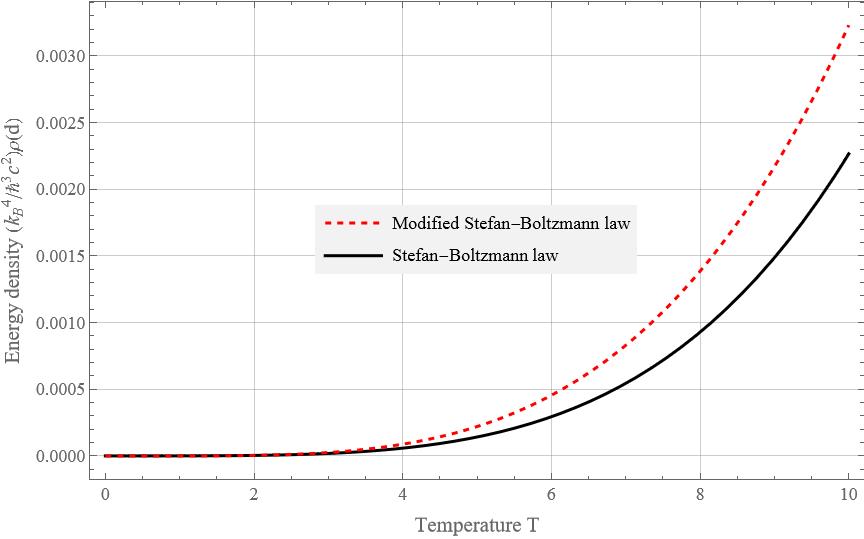}\label{}
		\caption{Comparison between the standard Stefan-Boltzmann law and the modified Stefan-Boltzmann law given by Eq. \eqref{73}. Here, $m=10$ is selected to exemplify the effect.}
		\label{fig1}
	\end{figure}

In the limit where the Podolsky mass becomes very large, $m \gg 1$, corresponding to $a \to 0$, Eq. (\ref{70}) reduces to
	\begin{align}
		\rho(T)=\frac{\pi^2T^4}{15}+\sum_{l_0=1}^{\infty}\left[9\left(\frac{m^3 T^5}{8\pi^3l_0^5}\right)^{1/2}+3\left(\frac{m^5 T^3}{8\pi^3l_0^3}\right)^{1/2}\right]e^{\frac{-ml_0}{T}},\label{73}
	\end{align}
where ${\cal T}^{00(11)}=\rho$. In this limit, an approximate expression for the Stefan-Boltzmann law is obtained, now including a correction arising from Podolsky's theory. Although the additional terms in brackets decay rapidly due to the exponential suppression, they still contribute non-negligibly to the total energy density. A comparison with the result presented in Ref. \cite{b2} shows that our expression reproduces the same leading term, but includes an additional contribution proportional to $m^{5/2}$. Figure \ref{fig1} illustrates the influence of the Podolski parameter on the Stefan-Boltzmann law where all physical constants have been restored to provide a more accurate description.
	
Starting from Eq. \eqref{70}, we examine the high-temperature limit, i.e., $\beta \to 0$. In this asymptotic regime, the Stefan-Boltzmann law takes the form
	\begin{align}
		\rho(T)=\frac{1}{\pi^2}\sum_{l_0=1}^{\infty}\left\{\frac{6}{\beta^4 l_0^4}+\frac{9}{\beta^4l_0^4}+\frac{3m^2}{2\beta^2l_0^2}\right\}=\frac{\pi^2T^4}{15}+\left[\frac{\pi^2T^4}{10}+\frac{m^2T^2}{4}\right],
	\end{align}
where the terms in brackets represent the corrections introduced by Podolsky's theory.
	
\subsection{Spatial confinement effects}\label{52}

\begin{figure}[!ht]
		\includegraphics[width=8cm]{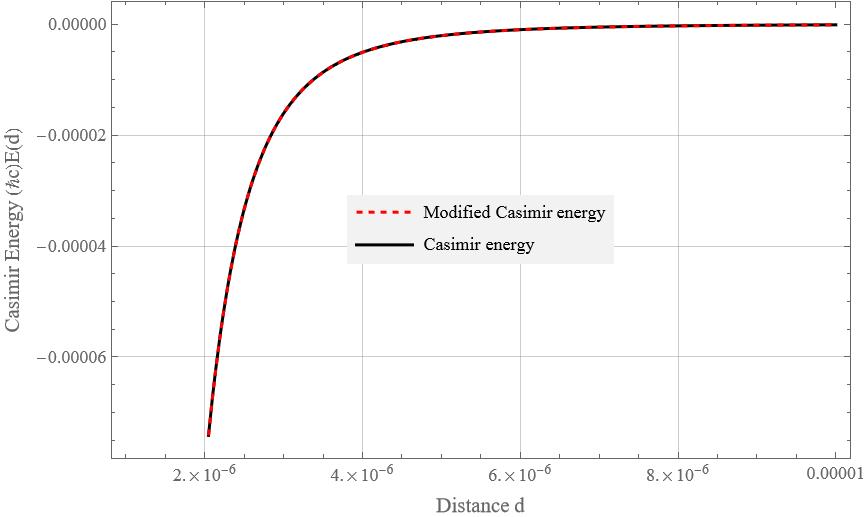}\label{}
		\caption{Comparison between the standard Casimir energy and the modified expression given by Eq. \eqref{79}. The correction introduced by Podolsky's theory is minimal, as it is exponentially suppressed by the factor $e^{-2md}$. For the parameter choice $m = 10$, the contribution from the Podolsky term rapidly decays with increasing distance $d$, resulting in a negligible deviation from the standard Casimir energy at large separations.}
		\label{fig2}
	\end{figure}

In this section, the Casimir effect is investigated by compactifying one of the spatial dimensions--specifically, the $z$-axis. This is implemented through the choice of the compactification parameter $\alpha = (0, 0, 0, i2d)$. Under this condition, the Bogoliubov transformation takes the form
	\begin{align}
		v^2(d)=\sum_{l_3=1}^{\infty}e^{-i2dk^3l_3},
	\end{align}
and the Green function becomes
	\begin{align}
		\overline{G}_0^P(x-x';d)=2\sum_{l_3=1}^{\infty}G_0^P(x-x'-2dl_3n_3).
	\end{align}
By selecting the component $\mu = \nu = 0$, which correspond to the Casimir energy, we obtain
	\begin{align}
		{\cal T}^{00(11)}(d)=-\frac{1}{\pi^2}\sum_{l_3=1}^{\infty}\left\{\frac{1}{8d^4l_3^4}+\frac{3m^2}{8d^2l_3^2}K_2(2mdl_3)\right\}.
	\end{align}
In the limit $m>>1$, the expression reduces to
	\begin{align}
		E(d)=-\frac{\pi^2}{720d^4}-\sum_{l_3=1}^{\infty}3\left(\frac{m^3}{256\pi^3d^5l_3^5}\right)^{1/2}e^{-2mdl_3}\label{79},
	\end{align}
where ${\cal T}^{00(11)}(d)=E(d)$. It is worth noting that the Casimir energy receives a small correction due to Podolsky's theory. Figure \ref{fig2} illustrates the behavior of the energy as a function of the separation distance, highlighting how the presence of the Podolsky parameter modifies the standard result.

To obtain the Casimir pressure, the component with $\mu = \nu = 3$ is considered, and as a result, the following expression is obtained	
\begin{align}
		{\cal T}^{33(11)}(d)=-\frac{1}{\pi^2}\sum_{l_3=1}^{\infty}\left\{\frac{3}{8d^4l_3^4}+\frac{9m^2}{8d^2l_3^2}K_2(2mdl_3)+\frac{3m^3}{4dl_3}K_1(2mdl_3)\right\}\label{80}.
	\end{align}
This represents the general expression for the Casimir pressure in terms of Podolsky's parameter. In the same asymptotic limit, $m>>1$, the pressure is found to be
	\begin{align}
		P(d)=-\frac{\pi^2}{240d^4}-\sum_{l_3=1}^{\infty}\left[9\left(\frac{m^3\pi}{256d^5l_3^5}\right)^{1/2}+3\left(\frac{m^5\pi}{64d^3l_3^3}\right)^{1/2}\right]e^{-2mdl_3}.
	\end{align}
Similar to the behavior observed for the Casimir energy, the pressure also acquires a correction due to Podolsky's electrodynamics, resulting in an overall increase in its magnitude.

\subsection{Thermal and spatial confinement effects}\label{53}

\begin{figure}[!ht]
		\includegraphics[width=8cm]{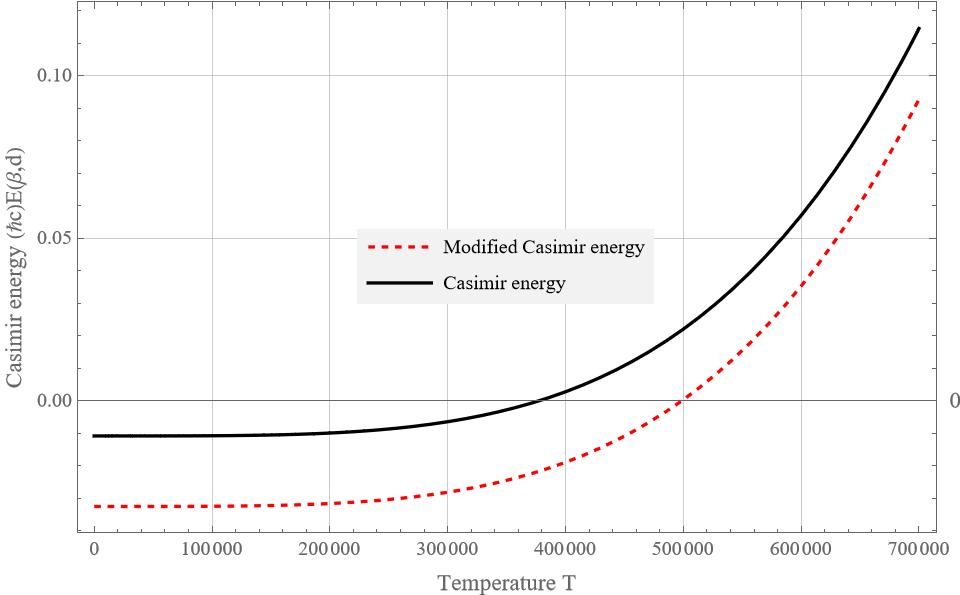}
		\caption{Comparison of the Casimir energy at finite temperature from Maxwell theory and its modification in Podolsky electrodynamics, as given by Eq. \eqref{85}, for $d=10^{-6}$ and $m=10$.}
		\label{fig3}
	\end{figure}

The main objective here is to describe the Casimir effect at finite temperature by compactifying both the time and one spatial dimension, such that $\alpha=(\beta,0,0,i2d)$. Accordingly, the Bogoliubov transformation is given by
	\begin{align}
		v^2(\beta,d)=\sum_{l_0=1}^{\infty}e^{-\beta k^0l_0}+\sum_{l_3=1}^{\infty}e^{-i2dk^3l_3}+\sum_{l_0,l_3=1}^{\infty}e^{-\beta k^0l_0-i2dk^3l_3}.
	\end{align}
The first two terms describe the thermal and size effects separately, as demonstrated in the previous subsections, while the third term accounts for the combined influence of thermal and size effects. The Green function associated with this third term is
	\begin{align}
		\overline{G}_0^P(x-x';\beta,d)=4\sum_{l_0,l_3=1}^{\infty}G_0^P(x-x'-i\beta l_0n_0-2dl_3n_3).
	\end{align}
To obtain the full expression for the Casimir energy, the component $\mu=\nu=0$ of the energy-momentum tensor is considered, yielding
	\begin{align}
		{\cal T}^{00(11)}(\beta,d)&=\frac{\pi^2T^4}{15}+\sum_{l_0=1}^{\infty}\left[9\left(\frac{m^3 T^5}{8\pi^3l_0^5}\right)^{1/2}+3\left(\frac{m^5 T^3}{8\pi^3l_0^3}\right)^{1/2}\right]e^{\frac{-ml_0}{T}}-\frac{\pi^2}{720d^4}\nonumber\\&-\sum_{l_3=1}^{\infty}3\left(\frac{m^3}{256\pi^3d^5l_3^5}\right)^{1/2}e^{-2mdl_3}+\frac{1}{\pi^2}\sum_{l_0,l_3=1}^{\infty}{\bigg\{}\frac{3(2\beta l_0)^2-(4dl_3)^2}{[(\beta l_0)^2+(2dl_3)^2]^3}\nonumber\\&+\frac{[(3\beta l_0)^2+6(2dl_3)^2]}{[(\beta l_0)^2+(2dl_3)^2]^3}m^2\beta^2l_0^2K_0(m\alpha)-\frac{3m^2(2dl_3)^4}{[(\beta l_0)^2+(2dl_3)^2]^3}K_0(m\alpha)\nonumber\\&+\frac{[2(3\beta l_0)^2-6(2dl_3)^2]}{[(\beta l_0)^2+(2dl_3)^2]^{5/2}}mK_1(m\alpha)+\frac{[3(\beta l_0)^2+3(2dl_3)^2]}{[(\beta l_0)^2+(2dl_3)^2]^{5/2}}m^3\beta^2l_0^2K_1(m\alpha){\bigg\}},
	\end{align}
where $\alpha=\sqrt{(\beta l_0)^2+(2dl_3)^2}$. This expression represents the complete Casimir energy at finite temperature, incorporating the modifications introduced by generalized electrodynamics. In the limit where $m>>1$, the above expression reduces to
	\begin{align}
		{\cal T}^{00(11)}(\beta,d)&=E(\beta,d)=\frac{\pi^2T^4}{15}+\sum_{l_0=1}^{\infty}\left[9\left(\frac{m^3 T^5}{8\pi^3l_0^5}\right)^{1/2}+3\left(\frac{m^5 T^3}{8\pi^3l_0^3}\right)^{1/2}\right]e^{\frac{-ml_0}{T}}\nonumber\\&-\frac{\pi^2}{720d^4}-\sum_{l_3=1}^{\infty}3\left(\frac{m^3}{256\pi^3d^5l_3^5}\right)^{1/2}e^{-2mdl_3}\nonumber\\&+\frac{1}{\pi^2}\sum_{l_0,l_3=1}^{\infty}{\bigg\{}\frac{3(2\beta l_0)^2-(4dl_3)^2}{[(\beta l_0)^2+(2dl_3)^2]^3}+{\bigg[}\frac{[(3\beta l_0)^2+6(2dl_3)^2]}{[(\beta l_0)^2+(2dl_3)^2]^{13/4}}m^2\beta^2l_0^2\nonumber\\&-\frac{3m^2(2dl_3)^4}{[(\beta l_0)^2+(2dl_3)^2]^{13/4}}+\frac{m[2(3\beta l_0)^2-6(2dl_3)^2]}{[(\beta l_0)^2+(2dl_3)^2]^{11/4}}\nonumber\\&+\frac{[3(\beta l_0)^2+3(2dl_3)^2]}{[(\beta l_0)^2+(2dl_3)^2]^{11/4}}m^3\beta^2l_0^2{\bigg]}\left(\frac{\pi}{2m}\right)^{1/2}e^{-m\sqrt{(\beta l_0)^2+(2dl_3)^2}}{\bigg\}}\label{85}.
	\end{align}
In a similar way, the Casimir pressure at finite temperature can be obtained by considering the component $\mu=\nu=3$. Figure \ref{fig3} presents a direct comparison between the Casimir energy in Podolsky electrodynamics and the standard result from Maxwell electrodynamics. It is observed that the presence of the Podolsky parameter noticeably influences the behavior of the energy, causing it to decrease at a slightly slower rate compared to the standard Casimir energy. However, as the temperature increases, both expressions tend to converge and approach similar values.

\section{Conclusions}\label{6}

In this work, a generalized form of electrodynamics proposed by B. Podolsky was investigated. This theory introduces a second-order derivative term in the electromagnetic field tensor along with a parameter $a$, which can be interpreted as generating a massive sector for the photon. Additionally, the Podolsky modification effectively acts as a high-frequency cutoff, addressing the divergences that arise in standard Maxwell electrodynamics \cite{p1,p2,p3}. As expected, the presence of this term leads to corrections in physical quantities such as the Stefan-Boltzmann law and the Casimir effect for the electromagnetic field.

The TFD formalism, a quantum field theory at finite temperature, was employed to describe thermal and spatial confinement effects through the analysis of different spacetime topologies. This approach is made possible by compactifying spacetime dimensions into a circular topology, effectively introducing thermal or boundary conditions \cite{tqft,ume}. In this work, three distinct topologies were considered: the compactification of the time axis, enabling the derivation of the Stefan-Boltzmann law; the compactification of the $z$-axis, associated with spatial confinement and the Casimir effect; and the simultaneous compactification of both axes, allowing for the analysis of the Casimir effect at finite temperature.

Therefore, it has been shown that the considered physical quantities are modified by the generalized electrodynamics. The Stefan-Boltzmann law is enhanced by the presence of the Podolsky parameter in both limits analyzed. Similarly, the Casimir energy and pressure at zero temperature are increased, leading to a stronger attractive force between the plates. At finite temperature, the behavior of the Casimir energy becomes more intricate due to the presence of multiple terms involving Bessel functions. However, under the condition $m \gg 1$, two regimes were explored. At low temperatures, a slight increase in energy was observed, with the correction depending solely on $\beta$. At high temperatures, the energy exhibits an exponential decay with temperature, driven by the influence of the Podolsky parameter.

\section*{Acknowledgments}

This work by A. F. S. is partially supported by National Council for Scientific and Technological
Development - CNPq project No. 312406/2023-1. L. H. A. R. Ferreira acknowledges CAPES for all the financial support provided.

\section*{Data Availability Statement}

No Data associated in the manuscript.

\section*{Conflicts of Interest}

No conflict of interests in this paper.
	

\global\long\def\link#1#2{\href{http://eudml.org/#1}{#2}}
 \global\long\def\doi#1#2{\href{http://dx.doi.org/#1}{#2}}
 \global\long\def\arXiv#1#2{\href{http://arxiv.org/abs/#1}{arXiv:#1 [#2]}}
 \global\long\def\arXivOld#1{\href{http://arxiv.org/abs/#1}{arXiv:#1}}


\end{document}